\documentclass[12pt]{article}
\pdfoutput=1

\usepackage{amsmath}
\usepackage{amsfonts}
\usepackage{amscd}
\usepackage{amsthm}
\usepackage{setspace}

\usepackage{graphicx}
\usepackage{authblk}
\usepackage{caption}
\usepackage{ytableau}
\usepackage{mathtools}

\def\Z{\mathbb{Z}}

\def\P{\mathbb{P}}

\def\til{\tilde}

\begin{document}

\begin{titlepage}

\begin{flushright}
KEK-TH-2168
\end{flushright}

\vskip 1cm

\begin{center}

{\bf \Large $\frac{1}{2}$Calabi--Yau 4-folds and four-dimensional F-theory on Calabi--Yau 4-folds with U(1) factors}

\vskip 1.2cm

Yusuke Kimura$^1$ 
\vskip 0.6cm
{\it $^1$KEK Theory Center, Institute of Particle and Nuclear Studies, KEK, \\ 1-1 Oho, Tsukuba, Ibaraki 305-0801, Japan}
\vskip 0.4cm
E-mail: kimurayu@post.kek.jp

\vskip 2cm
\abstract{In this study, four-dimensional $N=1$ F-theory models with multiple U(1) gauge group factors are constructed. A class of rational elliptic 4-folds, which we call as ``$\frac{1}{2}$Calabi--Yau 4-folds,'' is introduced, and we construct the elliptically fibered 4-folds by utilizing them. This yields a novel approach for building families of elliptically fibered Calabi--Yau 4-folds with positive Mordell--Weil ranks. The introduced $\frac{1}{2}$Calabi--Yau 4-folds possess the characteristic property wherein the sum of the ranks of the singularity type and the Mordell--Weil group is always equal to six. This interesting property enables us to construct the elliptically fibered Calabi--Yau 4-folds of various positive Mordell--Weil ranks. From one to six U(1) factors form in four-dimensional F-theory on the resulting Calabi--Yau 4-folds. We also propose the geometric condition on the base 3-fold of the built Calabi--Yau 4-folds that allows four-dimensional F-theory models that have heterotic duals to be distinguished from those that do not.}  

\end{center}
\end{titlepage}

\tableofcontents
\section{Introduction}
\par The aim of this study is to discuss the construction of four-dimensional (4D) $N=1$ F-theory models with multiple U(1) factors. To achieve the construction of such models, we introduced a certain family of rational elliptic 4-folds, which are referred to as ``$\frac{1}{2}$Calabi--Yau 4-folds'' in this study. The double covers of such rational elliptic 4-folds yields the elliptically fibered Calabi--Yau 4-folds. This construction approach yields families of Calabi--Yau 4-folds of various Mordell--Weil ranks. The F-theory compactifications on the resulting elliptically fibered Calabi--Yau 4-folds provide 4D $N=1$ models with U$(1)^n$, $n=1, \ldots, 6$ gauge groups. 
\par The U(1) gauge symmetry relates to the realization of the grand unified theory (GUT) because the presence of the U(1) gauge symmetry aids in explaining a few characteristic properties of the GUT such as the suppression of proton decay and mass hierarchy of quarks and leptons. We analyze the U(1) gauge group in the 4D $N=1$ F-theory. 
\par F-theory \cite{Vaf, MV1, MV2} is a formulation that extends type IIB superstrings to a nonperturbative regime. F-theory is compactified on spaces that admit a torus fibration. In the framework of F-theory, the axiodilaton in type IIB superstrings is identified with the modular parameters of elliptic curves as fibers of the torus fibration, thus enabling the axiodilaton to admit $SL_2(\Z)$ monodromy. 
\par The local model constructions \cite{DWmodel, BHV1, BHV2, DW} of F-theory model buildings have been mainly emphasized in recent studies. However, the global aspects of the geometry of the compactification space in F-theory should be studied to discuss issues pertaining to the early universe including inflation, and gravity. In this work, we analyze the structures of the compactification geometries from a global perspective. 

\vspace{5mm}

\par The U(1) gauge group arises in F-theory on an elliptic fibration when the elliptic fibration has a positive Mordell--Weil rank \cite{MV2}. Recent progress on F-theory models on elliptically fibered spaces admitting a global section can be found, for example, in \cite{MorrisonPark, MPW, BGK, BMPWsection, CKP, BGK1306, CGKP, CKP1307, CKPS, Mizoguchi1403, AL, EKY1410, LSW, CKPT, CGKPS, MP2, MPT1610, BMW2017, CL2017, BMW1706, KimuraMizoguchi, Kimura1802, LRW2018, TasilectWeigand, MizTani2018, TasilectCL, Kimura1810, CMPV1811, TT2019, Kimura1902, Kimura1903, EJ1905, LW1905, Kimura1910, CKS1910}. Recent studies of F-theory models possessing a U(1) gauge group were discussed, e.g., in \cite{MorrisonPark, BMPWsection, CKP, CGKP, BMPW2013, CKPS, MTsection, MT2014, KMOPR, BGKintfiber, GKK, MPT1610, CKPT, Kimura1802, TT2018, CLLO, CMPV1811, TT2019, Kimura1908, Kimura1910}.
\par The issue of flux \cite{BB, SVW, W96, GVW, DRS} arises in 4D F-theory models \footnote{F-theory models with four-form flux are recently discussed, e.g., in \cite{MSSN, CS, MSSN2, BCV, MS, KMW, GH, KMW2, IJMMP, KSN, CGK, BCV2, SNW, LLW1901}.}. The superpotential generated by the flux can modify the gauge groups and the matter spectra in 4D F-theory compactification, as discussed in \cite{MTsection}. We do not discuss the structure of the flux in this work. Furthermore, additional U(1) factors, which do not result from the rational sections in the 4D F-theory turning on fluxes, may arise, as reported in \cite{LW1905}. In this study, we focus on forming U(1) gauge groups in the 4D F-theory that originate from the Mordell--Weil group of the elliptic fibration. 

\vspace{5mm}

\par As mentioned previously, we introduced a class of rational elliptic 4-folds, which are referred to as ``$\frac{1}{2}$Calabi--Yau 4-folds'' in this study. These are obtained as a blow-up of six points on $\P^2\times \P^2$. These rational 4-folds naturally admit elliptic fibration, which is described in section \ref{sec3}. These rational elliptic 4-folds can be considered as a higher-dimensional generalization of rational elliptic surfaces \footnote{These surfaces are also referred to as $\frac{1}{2}$K3 surfaces. In this study, we refer to the elliptic surfaces obtained by blowing up the nine base points of a cubic pencil on the projective plane $\P^2$ as rational elliptic surfaces.}, and $\frac{1}{2}$Calabi--Yau 3-folds as introduced in \cite{Kimura1910}. As the double covers of such rational elliptic 4-folds provides elliptically fibered Calabi--Yau 4-folds, as discussed in section \ref{sec4.1}, we refer to them as $\frac{1}{2}$Calabi--Yau 4-folds in this study. A similar convention of the term was used in \cite{Kimura1910} to refer to a certain class of rational elliptic 3-folds. $\frac{1}{2}$Calabi--Yau 4-folds possess a characteristic property wherein the ranks of the Mordell--Weil group and the singularity type always add up to six. This property is analogous to those of rational elliptic surfaces and $\frac{1}{2}$Calabi--Yau 3-folds \cite{Kimura1910}, wherein the sums of the ranks add up to eight and seven, respectively. We build families of the elliptic Calabi--Yau 4-folds of various Mordell--Weil ranks by exploiting this property of $\frac{1}{2}$Calabi--Yau 4-folds; this is explained in section \ref{sec4.1}. 
\par The F-theory on the constructed Calabi--Yau 4-folds yields 4D $N=1$ models, with one to six U(1) factors. 

\vspace{5mm}

\par The base 3-folds of the elliptically fibered Calabi--Yau 4-folds obtained in this work are isomorphic to the Fano 3-folds of degree two \footnote{The degree of two means that $(-K)^3=2$ where $K$ denotes the canonical divisor class of the Fano 3-fold.}. We hypothesize that when the base Fano 3-fold admits a conic fibration \footnote{\cite{FMW, BJPS97, CD98, Andreas98, GT1204, AT1405, HT1506} discussed F-theory models on elliptic fibrations the base spaces of which admit a $\P^1$ fibration.}, the total elliptic Calabi--Yau 4-fold has a K3 fibration, which is compatible with the elliptic fibration \footnote{When an elliptically fibered Calabi--Yau 4-fold has a K3 fibration that is compatible with the elliptic fibration, the base 3-fold should admit a $\P^1$ fibration over a surface. However, the condition that the base 3-fold of a Calabi--Yau elliptic fibration is a conic fibration does not immediately imply that the total elliptic fibration has a K3 fibration that is compatible with the elliptic fibration. Our hypothesis here is limited to the situation in which the base 3-fold of an elliptic Calabi--Yau 4-fold is isomorphic to a Fano 3-fold of degree two.} based on natural reasoning. This is discussed in section \ref{sec5}. (When an elliptically fibered Calabi--Yau 4-fold has a compatible K3 fibration, the F-theory on the space has a dual heterotic string theory \footnote{The heterotic/F-theory duality \cite{Vaf, MV1, MV2, Sen, FMW} states that F-theory compactified on an elliptic K3 fibered Calabi--Yau $(n+1)$-fold over the base $B_{n-1}$ and heterotic string on an elliptically fibered Calabi--Yau $n$-fold over the same base $B_{n-1}$ are physically equivalent. F-theory/heterotic duality is strictly formulated when the stable degeneration limit \cite{FMW, AM} of a K3 fibration is considered for the F-theory. There have been recent studies on stable degenerations of F-theory/heterotic duality, such as \cite{BKW, BKL, AT1405, CGKPS, MizTan}.}.) In section \ref{sec5}, we also consider the effect of the condition that F-theory models possess heterotic duals on the non-Abelian gauge groups forming in the theory, by considering that the elliptic fibrations of the Calabi--Yau 4-folds also admit a compatible K3 fibration. Among the F-theory models constructed in this study, those possessing heterotic duals can receive constraints on the non-Abelian gauge groups arising on the 7-branes. Investigating the physical characteristics of the F-theory models possessing heterotic duals can be useful in understanding the structure of the string landscape. Analyzing the structure of the string landscape relates to the swampland conditions. Reviews of recent studies on the swampland criteria are given in \cite{BCV1711, Palti1903}. \cite{Vafa05, AMNV06, OV06} discussed the swampland.

\vspace{5mm}

\par The main results deduced in this study are summarized in section \ref{sec2}. The $\frac{1}{2}$Calabi--Yau 4-folds are introduced in section \ref{sec3}. The properties of these elliptic 4-folds related to the construction of F-theory models are also discussed. In section \ref{sec4.1}, elliptically fibered Calabi--Yau 4-folds with positive Mordell--Weil ranks are constructed by taking double covers of these 4-folds. The F-theory on the resulting Calabi--Yau 4-folds yields 4D $N=1$ models with multiple U(1) factors, as described in section \ref{sec4.2}. In section \ref{sec5}, the condition that elliptic fibrations of the Calabi--Yau 4-folds constructed in section \ref{sec4.1} admit a compatible K3 fibration is discussed in relation to the 4D F-theory/heterotic duality.

\section{Summary}
\label{sec2}
We summarize the results of this study in this section. As noted in the introduction, the main objective of this study is to systematically construct 4D $N=1$ F-theory models with multiple U(1) factors. In addition, as described in section \ref{sec5}, the geometric structure of the base 3-fold of the Calabi--Yau 4-folds constructed in this study can contain information that can be used to distinguish among the 4D F-theory models that possess heterotic duals and the models that do not possess heterotic duals. 
\par In section \ref{sec3}, we introduce the rational elliptic 4-folds that we refer to as $\frac{1}{2}$Calabi--Yau 4-folds. These 4-folds are obtained by blowing up six points on the product of the projective planes $\P^2\times\P^2$, and the resulting 4-folds have an elliptic fibration. The double covers of these elliptic 4-folds yield elliptically fibered Calabi--Yau 4-folds, as discussed in section \ref{sec4.1}. 
\par One of the most useful facts for this construction of Calabi--Yau 4-folds is that the $\frac{1}{2}$Calabi--Yau 4-folds $X$ should satisfy the following equation:
\begin{equation}
\label{equality in 2}
{\rm rk}\, {\rm MW} (X) + {\rm rk}\, {\rm ADE} (X)   =6.
\end{equation}
$MW(X)$ denotes the Mordell--Weil group of the $\frac{1}{2}$Calabi--Yau 4-fold $X$, and $ADE(X)$ is used to denote the singularity type of the $\frac{1}{2}$Calabi--Yau 4-fold $X$. This equation implies that once the rank of the singularity type of a $\frac{1}{2}$Calabi--Yau 4-fold has been determined, its Mordell--Weil rank can be automatically obtained. Generally, although it is considerably difficult to determine the Mordell--Weil rank of elliptically fibered manifolds by using the defining equation, owing to the equation (\ref{equality in 2}), determining the Mordell--Weil rank of $\frac{1}{2}$Calabi--Yau 4-folds is relatively easy. From the equation (\ref{equality in 2}), it can be deduced that the Mordell--Weil rank of $\frac{1}{2}$Calabi--Yau 4-fold ranges from zero to six. In this study, the members of the $\frac{1}{2}$Calabi--Yau 4-folds with Mordell--Weil ranks from one to six were considered. 
\par Calabi--Yau 4-folds constructed as double covers of the $\frac{1}{2}$Calabi--Yau 4-fold (the equation of which is given as (\ref{double cover CY in 4.1})) have Mordell--Weil rank greater than or equal to the original $\frac{1}{2}$Calabi--Yau 4-folds, as discussed in section \ref{sec4.1}. (The Mordell--Weil ranks of the original $\frac{1}{2}$Calabi--Yau 4-fold and the resulting Calabi--Yau 4-fold as a double cover are expected to be equal for the generic values of the parameters of the double cover.) Therefore, Calabi--Yau 4-folds of Mordell--Weil ranks (at least) one to six are obtained via the double covers of $\frac{1}{2}$Calabi--Yau 4-folds of Mordell--Weil ranks ranging from one to six. One to six U(1) factors form in 4D $N=1$ F-theory on the resulting Calabi--Yau 4-folds, depending on the Mordell--Weil rank of the constructed Calabi--Yau 4-folds. 
\par The $ADE$ singularity types of the original $\frac{1}{2}$Calabi--Yau 4-fold and the elliptically fibered Calabi--Yau 4-fold obtained as the double covers are identical, as demonstrated in section \ref{sec4.1}. Therefore, the non-Abelian gauge groups forming in F-theory on Calabi--Yau 4-folds as a double cover of $\frac{1}{2}$Calabi--Yau 4-fold can be deduced by determining the singularity type of the original $\frac{1}{2}$Calabi--Yau 4-fold. As discussed in section \ref{sec3}, the six points in $\P^2\times\P^2$ that are to be blown up to yield a $\frac{1}{2}$Calabi--Yau 4-fold are given as the intersection of four bidegree (1,1) hypersurfaces in $\P^2\times\P^2$. The singularity type of $\frac{1}{2}$Calabi--Yau 4-fold can, in principle, be determined by these (1,1) hypersurfaces.
\par As discussed in section \ref{sec4.1}, the base 3-fold of the elliptically fibered Calabi--Yau 4-folds constructed in this study is isomorphic to the Fano 3-fold of degree two. As described in section \ref{sec5}, we hypothesize that when this base 3-fold admits a conic fibration, the Calabi--Yau 4-fold has a K3 fibration. This hypothesis is natural, as explained in section \ref{sec5}. From the viewpoint of string theory, this yields a conjectural geometric condition determining when the 4D $N=1$ F-theory models obtained in this study possess heterotic duals. Furthermore, requiring the base space to have a conic fibration appears to impose constraints on the possible non-Abelian gauge groups forming on the 7-branes. This can imply that the 4D F-theory models that have heterotic duals receive constraints on the possible non-Abelian gauge groups, whereas those that do not have heterotic duals do not receive such constraints. If our geometric hypothesis is accurate, it appears to suggest that, at least for the 4D F-theory models on the constructed Calabi--Yau 4-folds in this study, the condition of the 4D F-theory models possessing heterotic duals translates to a geometric condition on the base 3-fold. This can be a useful clue to investigate the structure of the 4D $N=1$ F-theory landscape.

\section{$\frac{1}{2}$Calabi--Yau 4-folds}
\label{sec3}
In this section, we introduce a class of rational elliptic 4-folds that we refer to as $\frac{1}{2}$Calabi--Yau 4-folds. These 4-folds are used to build families of elliptically fibered Calabi--Yau 4-folds with various Mordell--Weil ranks, as mentioned in section \ref{sec4.1}. $\frac{1}{2}$Calabi--Yau 4-folds are obtained by blowing up six points on the product of projective planes $\P^2\times \P^2$. 
\par We consider four bidegree (1,1) hypersurfaces, $Q_1, Q_2, Q_3$, and $Q_4$, in $\P^2\times \P^2$. When one denotes the divisor class of a bidegree (1,1) hypersurface by $h_1+h_2$, then, as $(h_1+h_2)^4=6\, h_1^2h_2^2$, the four (1,1) hypersurfaces $Q_1, Q_2, Q_3, Q_4$ intersect at six points. These intersection points are the six points to be blown up to yield $\frac{1}{2}$Calabi--Yau 4-fold. The resulting 4-folds are rational by construction. Furthermore, they possess an elliptic fibration. This can be seen when the projection onto $\P^3$ is considered by taking the ratio
\begin{equation}
\label{projection onto P3 in 3}
[Q_1:Q_2: Q_3: Q_4].
\end{equation}
The fiber of this projection over the point $[a:b:c:d]$ in the base $\P^3$ is given as the complete intersection in $\P^2\times\P^2$:
\begin{eqnarray}
\label{complete intersection in 3}
bQ_1-aQ_2 & = 0 \\ \nonumber
cQ_2-bQ_3 & = 0 \\ \nonumber
dQ_3-cQ_4 & = 0. 
\end{eqnarray}
The complete intersection (\ref{complete intersection in 3}) is the intersection of three bidegree (1,1) hypersurfaces in $\P^2\times\P^2$, which is an elliptic curve; therefore, projection (\ref{projection onto P3 in 3}) yields an elliptic fibration over the base $\P^3$. 

\vspace{5mm}

\par The $\frac{1}{2}$Calabi--Yau 4-folds have a characteristic property that is very analogous to those of rational elliptic surfaces and $\frac{1}{2}$Calabi--Yau 3-folds introduced in \cite{Kimura1910}. The sum of the $ADE$ singularity rank and the Mordell--Weil rank of any $\frac{1}{2}$Calabi--Yau 4-fold $X$ is always six, independent of the complex structure. The equation (\ref{equality in 2}) in section \ref{sec2} expresses this property of the $\frac{1}{2}$Calabi--Yau 4-folds.  
\par Equation (\ref{equality in 2}) can be proved by applying an argument similar to that given in \cite{Kimura1910}. The outline of the proof is to note that the Picard number of any $\frac{1}{2}$Calabi--Yau 4-fold is 8 because the six-point blow-up of $\P^2\times\P^2$ increases the Picard number of $\P^2\times\P^2$ \footnote{The product $\P^2\times \P^2$ has the Picard number 2.} by 6; the Picard number gives the rank of the group generated by the divisors (modulo ``algebraic equivalence''). This group is generated by a zero-section and the smooth fiber class, and $\P^1$ components in the singular fibers not meeting the zero-section, and the group of the global sections. From this reasoning, the following equation \footnote{For elliptic surfaces possessing a global section an equation similar to the equation (\ref{equation divisors in 3}), which is known as the Shioda--Tate formula \cite{Shiodamodular, Tate1, Tate2} holds.} \footnote{The divisors in an elliptic fibration were discussed in \cite{Wazir, BM}.} holds:
\begin{equation}
\label{equation divisors in 3}
\rho(X)  =2+ {\rm rk}\, {\rm MW} (X) + {\rm rk}\, {\rm ADE} (X).
\end{equation}
Here, $\rho(X)$ denotes the Picard number of the $\frac{1}{2}$Calabi--Yau 4-fold, which is 8; thus, we obtain equation (\ref{equality in 2}). 

\vspace{5mm}

\par When the four bidegree (1,1) hypersurfaces $Q_1, Q_2, Q_3, Q_4$ are chosen generically, the configuration of the six intersection points is generic, and the resulting $\frac{1}{2}$Calabi--Yau 4-fold does not have an $ADE$ singularity. The Mordell--Weil rank of the $\frac{1}{2}$Calabi--Yau 4-fold is six owing to equation (\ref{equality in 2}) for this generic situation. 
\par When specific choices of the four bidegree (1,1) hypersurfaces $Q_1, Q_2, Q_3, Q_4$ are considered, the resulting $\frac{1}{2}$Calabi--Yau 4-folds have $ADE$ singularity types and the Mordell--Weil rank decreases and becomes less than six. 
\par The Mordell--Weil rank of $\frac{1}{2}$Calabi--Yau 4-folds ranges from zero to six. When $\frac{1}{2}$Calabi--Yau 4-folds develop singularity types of rank six such as $E_6$, $A_5A_1$, and $A_2^3$, they have the Mordell--Weil rank of zero. 
\par As shown in section \ref{sec4.1}, taking double covers of the $\frac{1}{2}$Calabi--Yau 4-folds yields elliptically fibered Calabi--Yau 4-folds, and the Mordell--Weil rank of the resulting Calabi--Yau 4-fold is greater than or equal to the Mordell--Weil rank of the original $\frac{1}{2}$Calabi--Yau 4-fold. As a result, double covers of $\frac{1}{2}$Calabi--Yau 4-folds having Mordell--Weil ranks from one to six give Calabi--Yau 4-folds of positive Mordell--Weil ranks. 

\section{Calabi--Yau 4-folds and 4D $N=1$ F-theory models with U(1) factors }
\label{sec4}

\subsection{Construction of elliptic Calabi--Yau 4-folds of positive Mordell--Weil ranks}
\label{sec4.1}
We construct elliptically fibered Calabi--Yau 4-folds of positive Mordell--Weil ranks by taking double covers of the $\frac{1}{2}$Calabi--Yau 4-folds introduced in section \ref{sec3}. As we see later in section \ref{sec4.2}, F-theory on the obtained Calabi--Yau 4-folds yields 4D models with multiple U(1) factors. 
\par We consider the double cover of of the $\frac{1}{2}$Calabi--Yau 4-folds of the following form:
\begin{equation}
\label{double cover CY in 4.1}
\tau^2 = F_6(Q_1, Q_2, Q_3, Q_4).
\end{equation}
In equation (\ref{double cover CY in 4.1}), we used $F_6$ to denote a degree-six polynomial in the bidegree (1,1) polynomials $Q_1, Q_2, Q_3, Q_4$ as variables \footnote{$F_6$ is a bidegree (6,6) polynomial in the coordinate variables of the product $\P^2\times \P^2$. The degree of the polynomial is selected to ensure that the double cover (\ref{double cover CY in 4.1}) satisfies the Calabi--Yau condition $K=0$.}. The double cover (\ref{double cover CY in 4.1}) ramified along the 3-fold $F_6(Q_1, Q_2, Q_3, Q_4)=0$ yields an elliptic Calabi--Yau 4-fold. 

\vspace{5mm}

\par As the base of the original $\frac{1}{2}$Calabi--Yau 4-fold was isomorphic to $\P^3$, by construction, the base 3-fold of the resulting elliptically fibered Calabi--Yau 4-fold (\ref{double cover CY in 4.1}) is a double cover of $\P^3$ ramified along a degree-six surface. This is known to be isomorphic to a Fano 3-fold of degree two by a well-known mathematical result. We denote the Fano 3-fold of degree two by $V_2$ in this study. 
\par As we discuss in section \ref{sec5}, we conjecture that when the base Fano 3-fold of degree two admits a conic fibration, the Calabi--Yau 4-fold (\ref{double cover CY in 4.1}) has a K3 fibration that is compatible with the elliptic fibration. This relates to 4D F-theory/heterotic duality. 

\vspace{5mm}

\par As taking a double cover (\ref{double cover CY in 4.1}) of the $\frac{1}{2}$Calabi--Yau 4-folds can be viewed as a base change, a global section of the original $\frac{1}{2}$Calabi--Yau 4-fold $X$ lifts to a global section of the Calabi--Yau 4-fold $Y$ (\ref{double cover CY in 4.1}); this situation is very analogous to those described in \cite{Kimura1802, Kimura1903, Kimura1910}. From this observation we learn that the Mordell--Weil group $MW(Y)$ of the Calabi--Yau 4-fold $Y$ (\ref{double cover CY in 4.1}) contains the Mordell--Weil group $MW(X)$ of the original $\frac{1}{2}$Calabi--Yau 4-fold $X$ as a subgroup. 
\par In particular, this indicates that the Mordell--Weil rank of the original $\frac{1}{2}$Calabi--Yau 4-fold $X$ is smaller than or equal to the Mordell--Weil rank of the Calabi--Yau 4-fold $Y$ (\ref{double cover CY in 4.1}). Therefore, we have 
\begin{equation}
\label{inequality of ranks in 4.1}
{\rm rk}\, {\rm MW} (X)\le {\rm rk}\, {\rm MW} (Y).
\end{equation}
\par We showed in section \ref{sec3} that the Mordell--Weil ranks of the $\frac{1}{2}$Calabi--Yau 4-folds range from zero to six owing to equation (\ref{equality in 2}). When the $\frac{1}{2}$Calabi--Yau 4-folds of Mordell--Weil ranks one to six are chosen and their double covers (\ref{double cover CY in 4.1}) are taken, owing to inequality (\ref{inequality of ranks in 4.1}), Calabi--Yau 4-folds of positive Mordell--Weil ranks are obtained. 
\par Furthermore, it is expected that the equality holds in (\ref{inequality of ranks in 4.1}) for generic values of the parameters of taking the double cover (\ref{double cover CY in 4.1}). Utilizing an argument similar to that given in \cite{Kimura1802, Kimura1910} leads to this expectation by considering a limit at which a Calabi--Yau 4-fold (\ref{double cover CY in 4.1}) splits into a pair of $\frac{1}{2}$Calabi--Yau 4-folds \footnote{A deformation of the Calabi--Yau 4-fold (\ref{double cover CY in 4.1}), similar to the processes discussed in \cite{KRES, Kimura1910}, can be considered as
\begin{equation}
\label{deformation of double cover in 4.1}
\tau^2 = F_6(Q_1, Q_2, Q_3, Q_4) +\lambda \, G_3(Q_1, Q_2, Q_3, Q_4)^2,
\end{equation}
where $\lambda$ denotes the parameter of a deformation. We have used $G_3$ to denote the degree-three polynomial in the (1,1) polynomials $Q_1, Q_2, Q_3, Q_4$, given by $G_3=(a_1\, Q_1+a_2\, Q_2+a_3\, Q_3)(a_4\, Q_1+a_5\, Q_2+a_6\, Q_4)(a_7\, Q_2+a_8\, Q_3+a_9\, Q_4)$. (Here, $a_i$, $i=1, \ldots, 9$ are constant.) Setting $\lambda=0$ gives the Calabi--Yau double cover (\ref{double cover CY in 4.1}). When the limit at which $\lambda$ goes to $\infty$ is considered, the Calabi--Yau 4-fold (\ref{deformation of double cover in 4.1}) is split into a pair of $\frac{1}{2}$Calabi--Yau 4-folds as $\tau= \pm G_3(Q_1, Q_2, Q_3, Q_4)$.}, and a global section of the Calabi--Yau 4-fold also splits to yield global sections of the two $\frac{1}{2}$Calabi--Yau 4-folds into which the Calabi--Yau 4-fold is split. 

\vspace{5mm}

\par The original $\frac{1}{2}$Calabi--Yau 4-fold and the Calabi--Yau 4-fold obtained as its double cover (\ref{double cover CY in 4.1}) have identical singularity types. We expand the discriminant of $\frac{1}{2}$Calabi--Yau 4-fold into irreducible factors as:
\begin{equation}
\Delta (X) = \prod p_i^{m_i}.
\end{equation}
We have used $\Delta(X)$ to denote the discriminant of the original $\frac{1}{2}$Calabi--Yau 4-fold $X$. The operation of taking double cover (\ref{double cover CY in 4.1}) corresponds to a base change, and the polynomial $p_i$ is replaced with $\til{p}_i$ under this operation. $\til{p}_i$ is obtained by plugging polynomials that correspond to the base change into the coordinate variables of $\P^3$ of the polynomial $p_i$. Similar to the situation as discussed in \cite{Kimura1910}, because the base of the Calabi--Yau 4-folds and the $\frac{1}{2}$Calabi--Yau 4-folds are 3-folds, unlike the case where the base is $\P^1$ \cite{KRES, Kimura1802, Kimura1903}, the polynomial $\til{p}_i$ obtained via the base change remains irreducible, and 
\begin{equation}
\Delta (Y) = \prod \til{p}_i^{m_i}.
\end{equation}
yields irreducible decomposition of the discriminant of the Calabi--Yau 4-fold $Y$ (\ref{double cover CY in 4.1}). $\Delta (Y)$ is used to denote the discriminant of the Calabi--Yau 4-fold $Y$ obtained as the double cover (\ref{double cover CY in 4.1}) of $X$. Thus, from an argument similar to that given in \cite{Kimura1910}, it is evident that the singularity type of the Calabi--Yau 4-fold (\ref{double cover CY in 4.1}) is identical to that of the original $\frac{1}{2}$Calabi--Yau 4-fold.

\subsection{4D F-theory models with U(1) factors}
\label{sec4.2}
We constructed elliptically fibered Calabi--Yau 4-folds of positive Mordell--Weil ranks in section \ref{sec4.1}. When the $\frac{1}{2}$Calabi--Yau 4-fold has Mordell--Weil rank $n$, the resulting Calabi--Yau 4-fold obtained by taking double cover (\ref{double cover CY in 4.1}) has the Mordell--Weil rank of $\ge n$, $n=1, \cdots, 6$, owing to the inequality (\ref{inequality of ranks in 4.1}). As stated previously, it is expected that their Mordell--Weil ranks are actually equal for the generic parameters of the double cover (\ref{double cover CY in 4.1}). U$(1)^n$ forms in F-theory on the Calabi--Yau 4-fold obtained as double cover (\ref{double cover CY in 4.1}) of $\frac{1}{2}$Calabi--Yau 4-fold with the Mordell--Weil rank $n$, $n=1, \ldots, 6$. 
\par When a $\frac{1}{2}$Calabi--Yau 4-fold has Mordell--Weil rank strictly less than six, it must also have an $ADE$ singularity to satisfy the relation (\ref{equality in 2}). Calabi--Yau 4-fold as the double cover (\ref{double cover CY in 4.1}) of this $\frac{1}{2}$Calabi--Yau 4-fold has an identical $ADE$ singularity type, and a non-Abelian gauge group arises on the 7-branes in F-theory on the resulting Calabi--Yau 4-fold in this situation. 
\par When the four bidegree (1,1) hypersurfaces $Q_1, Q_2, Q_3, Q_4$ are chosen to be generic, the $\frac{1}{2}$Calabi--Yau 4-fold obtained by blowing up the base points of these four hypersurfaces in $\P^2\times\P^2$ does not have an $ADE$ singularity as we noted previously; therefore, it has Mordell--Weil rank six owing to equation (\ref{equality in 2}). An elliptically fibered Calabi--Yau 4-fold obtained as double cover (\ref{double cover CY in 4.1}) of this $\frac{1}{2}$Calabi--Yau 4-fold has Mordell--Weil rank of (at least) six, and the resulting Calabi--Yau 4-fold does not have an $ADE$ singularity. F-theory on this Calabi--Yau 4-fold yields a 4D $N=1$ model with U$(1)^6$ gauge group, and this model does not have a non-Abelian gauge group factor.

\section{K3 fibration that is compatible with elliptic fibration, and relation with heterotic duals}
\label{sec5}
\begin{figure}
\begin{center}
\includegraphics[height=10cm, bb=0 0 960 540]{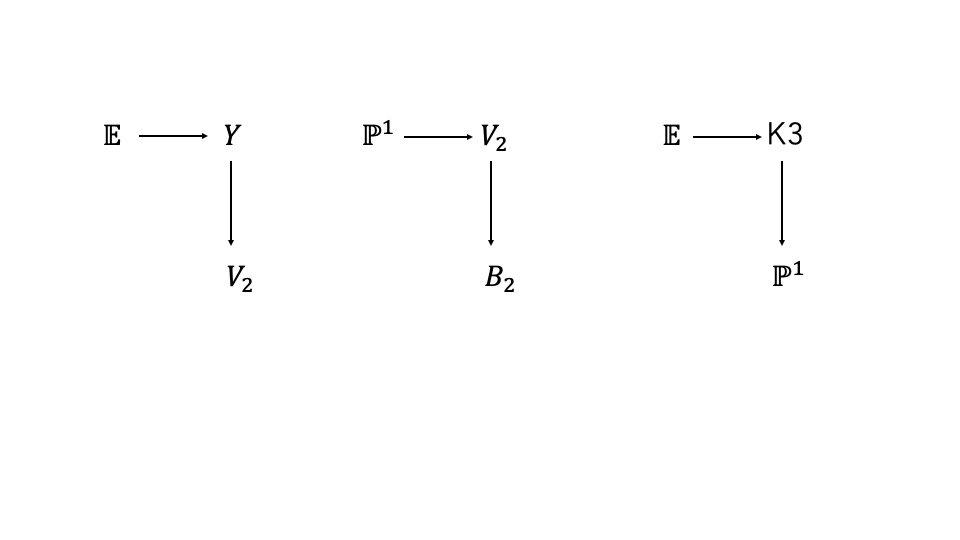}
\caption{\label{imagesofbundlesin5}Elliptically fibered Calabi--Yau 4-fold $Y$ with the base Fano 3-fold of degree two $V_2$, and Fano 3-fold of degree two $V_2$ as a conic fibration with the base surface $B_2$ when the Fano 3-fold of degree two admits a conic fibration. Elliptic fibration over base $\P^1$ yields a K3 surface when the total space of the fibration is required to satisfy the Calabi--Yau condition.}
\end{center}
\end{figure}

We learned in section \ref{sec4.1} that the elliptically fibered Calabi--Yau 4-folds constructed in section \ref{sec4.1} have base 3-fold isomorphic to a Fano 3-fold of degree two, $V_2$. We conjecture that when this base Fano 3-fold of degree two $V_2$ admits a conic fibration, the Calabi--Yau 4-fold admits a K3 fibration that is compatible with the elliptic fibration. This is very natural to expect, and this expectation is based on the following observation: when the base 3-fold of a Calabi--Yau 4-fold has a conic fibration, the base 3-fold is a $\P^1$ fibration over the base surface. The total Calabi--Yau 4-fold, then, is a fibration over the base surface, and the fiber of this fibration is an elliptic fibration over $\P^1$. Elliptic fibration over $\P^1$ (with the condition that the total space of this fibration satisfies the Calabi--Yau condition) yields a K3 surface. Thus, it is natural to expect that when the base degree-two Fano 3-fold is a conic fibration over a surface, the total Calabi--Yau 4-fold is quite likely to be a K3 fibration over the base surface, and this K3 fibration is compatible with the elliptic fibration. Figure \ref{imagesofbundlesin5} shows images of these.

\vspace{5mm}

\par Among the 4D $N=1$ F-theory models constructed in this study, those having heterotic duals can have some additional constraints on the possible non-Abelian gauge groups that can form on the 7-branes.
\par Non-Abelian gauge groups arising on the 7-branes in F-theory are determined by the types of the singular fibers \footnote{The types of the singular fibers of the elliptic surfaces were classified by Kodaira in \cite{Kod1, Kod2}. Techniques to determine the singular fiber types of elliptic surfaces can be found in \cite{Ner, Tate}.} and by whether they are split/non-split/semisplit \cite{BIKMSV}, and these are determined by the Weierstrass coefficients of the elliptic fibration of the compactification space. It is natural to expect that requiring the base 3-fold to have an additional structure such as a conic fibration imposes some constraints on the Weierstrass coefficients of the elliptic fibration of the Calabi--Yau 4-fold; therefore, requiring the base 3-fold to have an additional structure such that the total elliptic fibration also admits a compatible K3 fibration is likely to restrict the types of singular fibers. Thus, from the physical viewpoint, it is likely to restrict the possible non-Abelian gauge groups forming on the 7-branes.
\par It would be interesting to consider the physical constraints on the possible non-Abelian gauge groups on the 7-branes when a condition is imposed on the elliptically fibered Calabi--Yau 4-folds to also admit a K3 fibration that is compatible with elliptic fibration, and studying this physical effect is a likely direction for future work. This can help in analyzing the structure of the 4D $N=1$ F-theory landscape. One can compare the non-Abelian gauge groups forming in 4D F-theory models that have heterotic duals with those that do not have heterotic duals. 
\par The rank of the Mordell--Weil group corresponds to the number of ways one can embed the base space into the total elliptic fibration \cite{MV2}. As requiring the base 3-fold to have an additional structure such that the elliptic fibration also has a K3 fibration, and we conjectured that a conic fibration yields an example of this additional structure when the base 3-fold is isomorphic to the Fano 3-fold of degree two, does not appear to affect the way the base 3-fold is embedded into the total elliptic fibration, we expect that whether the constructed elliptically fibered Calabi--Yau 4-folds admit a compatible K3 fibration does not place a strong effect on the number of U(1) factors forming in F-theory compactifications. Thus, it is expected, at least for the 4D F-theory models on the constructed Calabi--Yau 4-folds in this study, that whether a 4D F-theory model has heterotic dual does not have a significant effect on the number of U(1) factors forming in theory. We only discussed forming U(1) factors in 4D F-theory that originate from the Mordell--Weil group in this study, as we noted in the introduction.

\section{Conclusions and open problems}
\label{sec6}
Here, we constructed elliptically fibered Calabi--Yau 4-folds of positive Mordell--Weil ranks. F-theory on the resulting Calabi--Yau spaces yields 4D $N=1$ models with one to six U(1) factors \footnote{4D $N=1$ F-theory models without a U(1) gauge group can be found, for example, in \cite{KCY4, Kdisc}.}. 
\par Investigating explicit constructions of $\frac{1}{2}$Calabi--Yau 4-folds with $ADE$ singularity types is a likely direction of future study. Taking double covers of such elliptic 4-folds yields examples of elliptic Calabi--Yau 4-folds with $ADE$ singularity types identical to the original $\frac{1}{2}$Calabi--Yau 4-folds. F-theory on the resulting Calabi--Yau spaces yields explicit examples of 4D $N=1$ theories with multiple U(1) factors and non-Abelian gauge groups.
\par The $\frac{1}{2}$Calabi--Yau 4-folds built in section \ref{sec3} do not have $E_7$ and $E_8$ singularities owing to equation (\ref{equality in 2}) because $E_7$ and $E_8$ singularities have ranks greater than six. Therefore, in this study, our construction of Calabi--Yau 4-folds by taking the double cover (\ref{double cover CY in 4.1}) does not, at least directly, provide Calabi--Yau 4-folds with $E_7$ or $E_8$ singularity. Determining whether elliptically fibered Calabi--Yau 4-folds over the base Fano 3-fold of degree two, $V_2$, with $E_7$ and $E_8$ singularities exist is an open problem. 
\par When a Calabi--Yau 4-fold admits a K3 fibration (that is compatible with the elliptic fibration), considering the heterotic dual can aid in analyzing this problem.
\par Deducing the Weierstrass equations of the Calabi--Yau 4-folds constructed in this work from the equations of four bidegree (1,1) hypersurfaces in $\P^2\times\P^2$ can also be a likely target for future studies. Because the equations of four bidegree (1,1) hypersurfaces determine the defining equation of $\frac{1}{2}$Calabi--Yau 4-fold, the Weierstrass equation of Calabi--Yau 4-fold as double cover (\ref{double cover CY in 4.1}) of the $\frac{1}{2}$Calabi--Yau 4-fold can in principle be obtained from the equations of four bidegree (1,1) hypersurfaces, $Q_1, Q_2, Q_3, Q_4$. The Weierstrass equations of the Calabi--Yau 4-folds help to obtain the Yukawa couplings of the matter fields arising in 4D F-theory.

\section*{Acknowledgments}

We would like to thank Shun'ya Mizoguchi and Shigeru Mukai for discussions.

\end{document}